# Observation of an electrically tunable band gap in trilayer graphene

Chun Hung Lui[1], Zhiqiang Li[1], Kin Fai Mak[1], Emmanuele Cappelluti[2], and Tony F. Heinz[1*]

[1] Departments of Physics and Electrical Engineering, Columbia University, 538 West 120th Street, New York, NY 10027, USA

[2] Institute for Complex Systems (ISC), CNR, v. dei Taurini 19, 00185 Rome, Italy

*corresponding author (email: tony.heinz@columbia.edu)

**A striking feature of bilayer graphene is the induction of a significant band gap in the electronic states by the application of a perpendicular electric field[1-6]. Thicker graphene layers are also highly attractive materials. The ability to produce a band gap in these systems is of great fundamental and practical interest[7-23]. Both experimental[10] and theoretical[14,21,22] investigations of graphene trilayers with the typical ABA layer stacking have, however, revealed the lack of any appreciable induced gap. Here we contrast this behavior with that exhibited by graphene trilayers with ABC crystallographic stacking. The symmetry of this structure is similar to that of AB stacked graphene bilayers and, as shown by infrared conductivity measurements, permits a large band gap to be formed by an applied electric field. Our results demonstrate the critical and hitherto neglected role of the crystallographic stacking sequence on the induction of a band gap in few-layer graphene.**

Producing a controlled and tunable band gap in graphene is a topic of central importance[1-6,14,21-25]. In addition to the intrinsic interest of altering the fundamental electronic properties of materials, the availability of an adjustable band gap opens up the possibility of a much wider range of applications for graphene in electronics and photonics. Both single- and few-layer graphene in their unperturbed state lack a band gap[13,16]. However, few-layer graphene materials under the application of a symmetry-lowering perpendicular electric field may exhibit an induced gap[1,2,4-6,14,21-24]. In this regard, trilayer graphene is an attractive material system. Unlike bilayer graphene, however, trilayers, which typically exhibit Bernal (ABA) stacking order[19] and the associated mirror symmetry (figure 1a), have been shown both theoretically[14,21,22] and experimentally[10] not to support the induction of a significant band gap when subjected to a perpendicular electric field. As discussed below, this behavior follows from the mirror symmetry of the unperturbed ABA trilayer[14]. Recent research[26] has, however, reported the existence of a new type of trilayer graphene, one with ABC (rhombohedral) stacking order between the graphene sheets[13,14,18,24] (figure 1b). This crystal structure, like that of the bilayer possesses inversion symmetry, but lacks mirror symmetry (figure 1b). The electronic structure of the ABC trilayer[16,24] is accordingly more similar to that of the AB-stacked bilayer graphene. In particular, the undoped ABC trilayer has only two-fold degeneracy[16] at the Fermi energy, like the graphene bilayer, rather than the four-fold degeneracy found in the ABA trilayer[16,19]. The two-fold degeneracy in the ABC trilayer band structure can be readily lifted by imposing different potentials on the top and bottom graphene layers by an applied electric field, which leads to the opening of a band gap[14,18,23,24]. While theory has predicted the induction of a large band gap for ABC trilayer graphene, experimental confirmation has been lacking.

In this paper, we report an experimental and theoretical study of the electronic response of trilayer graphene, both of ABA and ABC stacking order, to perpendicular electric fields as strong as ~0.2 V/nm. Our results show direct spectroscopic signatures of the induction of a tunable band gap of as much as ~120 meV in ABC trilayer graphene. Such a band gap is not observable in ABA trilayers under the same electric field. We analyze these results by considering the implications of the different crystal structure and interlayer coupling in ABA and ABC stacked trilayers.

We investigated graphene trilayer samples exfoliated from kish graphite on SiO$_2$/Si substrates. The sample thickness and stacking order were first determined by infrared[8,9,27] and Raman[26] spectroscopy. In our measurements, we made use of an electrolyte top gate[4,28] (figure 2a) to induce high doping densities and electric fields in the samples. The resultant change of the band structure is probed by infrared conductivity measurements[4,8,9,27] (see Methods).

We have measured IR sheet conductivity $\sigma(\hbar\omega)$ of ABA and ABC trilayer graphene samples at different gate voltages $V_g$ (figure 2). At the charge neutrality point ($V_g = V_{CN}$), the ABC spectrum shows a single absorption peak at $\hbar\omega = 0.35$ eV (figure 2b), and the ABA spectrum exhibits two peaks at 0.52 and 0.585 eV (figure 2f). These transitions reflect the distinct nature of the interlayer interactions and low-energy band structure for the two types of crystal structures (figure 2e,i). The energies of the absorption peaks in ABC and ABA trilayer correspond approximately to $\gamma_1$ and $\sqrt{2}\gamma_1$, respectively, where $\gamma_1$~0.37 eV is the nearest-neighbor *interlayer* coupling strength. The factor of $\sqrt{2}$ arises from the mirror symmetry in ABA trilayer[8,16], where the atoms in the middle layer are coupled symmetrically with atoms in both the bottom and top layers (figure 1a,c). We note that the two slightly different transition energies of 0.52 and 0.585 eV in ABA trilayer correspond, respectively, to hole and electron transitions[29]. (See Supplementary Information for more detailed analysis of the electron-hole asymmetry in ABA trilayer).

As we increase the gate bias for the ABC trilayer, the main peak splits into two distinct features (*P1* and *P2* in figure 2b) that shift in opposite directions and broaden. This behavior is a clear signature of the induction of a band gap. Corresponding effects are also observed when a negative gate voltage is applied to produce hole doping (as described in the Supplementary Information). Figure 2e shows the evolution of the electronic structure of ABC trilayer graphene under an applied electric field according to a tight-binding (TB) calculation that includes the dominant intralayer $\gamma_0$ and interlayer $\gamma_1$ couplings. The unperturbed ABC trilayer (green line) has three valence and conduction bands near the *K*-point in the Brillouin zone. The two low-energy bands touch one another at the *K*-point, while the other bands are separated by $\gamma_1$~370 meV. With the application of a strong electric field, a gap develops between the low-energy valence band and conduction band (red line). The observed absorption peaks *P1* and *P2* are readily understood as arising from the transitions indicated as *1* and *2* in the modified band structure. The difference between *P1* and *P2* hence reflects the size of the band gap, which reaches ~120 meV at the largest applied gate voltage of 1.2V.

For the ABA trilayer, as we increase the gate bias, the amplitude of the transition at 0.585 eV grows and the peak position red shifts, while the low-energy peak at 0.520 eV disappears (figure 2f). A similar effect was observed for negative gate biases and hole doping (see Supplementary Information). Apart from state-filling effects that reflect the increase of Fermi level under gating, there is no evidence of the emergence of additional peaks associated with the creation of a band gap. We estimate from the broadening of the absorption peak that an induced band gap, if it exists, should not exceed 30 meV at the highest gating voltage of 0.9 V.

The above observations can be understood within a framework of the TB description, with a self-consistent scheme[22] to take into account the gate-induced electric field across the graphene layers (see Methods). For the ABC case, we consider only the dominant coupling terms of $\gamma_0$ and $\gamma_1$. Carrying out TB calculation with a full set of coupling parameters did not yield significantly different predictions. To obtain the best fit to the data, we used a value for the interlayer coupling of $\gamma_1 = 377$ meV and assumed a capacitance of the electrolyte top gate of $C_g = 1.3$ μF cm$^{-2}$. The predicted band gap, $E_g$, and the energy gap at *K*-point, $\Delta E_k$, agree well with the band gap extracted from experiment (figure 3). For more detailed and direct comparison, we calculated the expected IR conductivity spectra by means of Kubo formula (figure 2c). These simulations clearly reproduce the main features of the experimental spectra (figure 2b). We also show for comparison the predicted conductivity under the neglect of any induced modification of the electronic structure or band gap opening (figure 2d), including only the effect of state filling on the optical transitions. The resulting behavior is completely inconsistent with experiment.

In the case of the ABA trilayer structure, we include in the TB simulations parameters that describe the observed electron-hole asymmetry. In particular, we use $\delta = 37$ meV as the average on-site energy difference between atomic sites A1, B2, A3 and B1, A2, B3 (figure 1c) and $v_4 \equiv \gamma_4/\gamma_0 = 0.05$ to describe the next-nearest-neighbor interlayer coupling strength. We found reasonable agreement between the experiment and the simulated $\sigma(\hbar\omega)$ spectra by the Kubo formula (figure 2g) with similar values for the other parameters in the model as for ABC stacking ($\gamma_1=371$ meV and $C_g = 0.8$ μFcm$^{-2}$). For comparison, we also show the predicted $\sigma(\hbar\omega)$ spectra under the neglect of any induced modification of the band structure (figure 2d). The resultant spectra are rather similar to the previous simulations (figure 2c). This conclusion is consistent with a predicted band structure for the ABA trilayer that changes little under the applied electric field (figure 2i).

As this analysis shows, the induction of a gap in graphene trilayers is completely different for ABA- and ABC-stacked materials. For applied electric fields of similar strength, the ABC trilayer shows a sizable band gap of ~120 meV, while the ABA trilayer does not exhibit any signature of band-gap opening. The different behaviors can be understood within a TB model using just the dominant intra- and interlayer parameters of $\gamma_0$ and $\gamma_1$ (Figure 1c,d). At the *K*-point of the Brillouin zone, the effective *intralayer* coupling vanishes[14]. The states of ABC trilayer can hence be represented by two dimers with finite energies ($\pm\gamma_1$) and two monomers with zero energy (blue and yellow atoms in figure 1d, respectively). The application of a perpendicular electric field induces different potentials at the bottom and top layers. This lifts the degeneracy of the two corresponding monomer states (A1 and B3) and induces a band gap[14]. On the other hand, the electronic states at the *K*-point in the ABA trilayer system are represented by a trimer and three monomers (blue and yellow atoms in figure 1c, respectively). The trimer has a non-bonding state that forms a four-fold degenerate zero-energy level with the monomers. While a vertical electric field can lift the degeneracy of the two monomer states on the bottom and top layers (A1 and A3), it has no appreciable influence on the monomer state on the middle layer (B2) and the non-bonding trimer state. The presence of this remaining degeneracy precludes the induction of a band gap in ABA trilayers[14].

It is informative to compare our results with the behavior found in bilayer graphene under the influence of an applied electric field[3,4]. For the ABC-stacked trilayer, we have observed an induced band gap of 120 meV for an applied electric field of ~0.2 V/nm. Induction of a comparable gap in bilayer graphene is achieved for an applied field of 0.4 – 0.5 V/nm[3,4]. The increased sensitivity to the applied field for the trilayer sample is expected because the size of the induced band gap for a given field increases with layer thickness. In particular, for the same (moderate) applied field, the band gap in the thicker ABC trilayer should be approximately twice as large as in the (AB) bilayer[18], in agreement with our observations. In addition, for applications involving a material with tunable infrared properties, we found that the infrared peaks in ABC trilayers are much sharper than those observed in bilayers[4] because of the higher-order van Hove singularity in ABC trilayer band structure[16]. These better defined features favor applications of trilayers for applications requiring a tunable change in IR absorption. More generally, our work suggests that a tunable band gap can be induced in thicker graphene samples with ABC (rhombohedral) stacking order[9,16,17,20], thus providing a still broader class of materials with a tunable band gap.

## Methods

**Sample preparation and characterization** Graphene trilayer samples were prepared by mechanical exfoliation of kish graphite (Toshiba) on silicon substrates coated with a 300-nm oxide layer. The sample thickness and stacking order are characterized by means of infrared spectroscopy[9,26]. These measurements were performed using the National Synchrotron Light Source at Brookhaven National Laboratory (U12IR beam line). For a more detailed analysis of the spatial variation of the sample, we relied on scanning Raman spectroscopy[26]. Using the signature of the stacking-order in the 2D Raman feature, we could visualize the spatial distribution of the ABA and ABC stacking domains in trilayer samples. We found ~60% of trilayer samples were of purely ABA stacking order, while the rest exhibited mixed ABA-ABC stacking orders. For our investigations, we chose for device fabrications those samples showing either pure ABA stacking or large (>200 µm$^2$) homogeneous domains of ABC stacking.

**Device Fabrication** The overall gated device structure involved top gating of the graphene trilayer sample with a polymer electrolyte gate (figure 2a). Electrical contacts to the graphene samples were formed using electron-beam lithography and electron-beam evaporation of Au films of 50-nm thickness. The polymer electrolyte (poly(ethylene oxide): LiClO$_4$, 8:1, dissolved in methanol)[4,28] was then cast onto the sample and dried at 110 ºC in ambient. A large Au electrode, deposited within 100 µm of the graphene samples, provided electrical contact to the transparent polymer gate. The capacitance of such top gates was typically ~1.0 µFcm$^{-2}$ and thus allowed us to induce charge densities of ~10$^{13}$ cm$^{-2}$.

**Determination of the optical conductivity** We measured the infrared transmission spectrum of the gated trilayers by normalizing the sample spectrum with that from the bare substrate. We then extracted the real part of the optical sheet conductivity ($\sigma$) in the spectral range of 0.2-1.0 eV from the transmission spectra by solving the optical problem for a thin film on the SiO$_2$(300-nm)/Si substrate. In our calculation, we neglect the interference from the sample/PEO interface and consider only the much stronger reflection from SiO$_2$/Si interface. We also neglect the contribution of the imaginary part of the optical conductivity. The above simplifications are estimated to induce 10% errors in $\sigma$, mainly in the spectral range below 0.3 eV, and have negligible influence on the spectral positions of the peaks in $\sigma$.

**Theoretical simulation of the optical conductivity** We use the self-consistent approach of Avetisyan *et al*[22] to calculate the charge density at each layer of the graphene trilayers for different total charge density. In the calculation, we consider only the dominant $\gamma_0$ and $\gamma_1$ couplings in ABC trilayer TB Hamiltonian and $\gamma_0$, $\gamma_1$, $\gamma_3$, and $\delta$ in ABA trilayer TB Hamiltonian. We use the dielectric constant of bulk graphite ($\kappa=2.4$) in the calculation. With the self-consistent charge distribution, we simulate the optical conductivity by using the Kubo formula with a broadening parameter of 10 meV.

**Acknowledgment** We thank G. L. Carr and R. Smith for technical support in the infrared measurement at Brookhaven National Laboratory, D. Efetov for support in device fabrication and A. A. Avetisyan, B. Partoens, F. M. Peeter, M. Koshino, and Y. L. Li for discussions. The authors at Columbia University acknowledge support from the Office of Naval Research under the MURI program, from the National Science Foundation under Grant CHE-0117752, and from the New York State Office of Science, Technology, and Academic Research (NYSTAR). E.C. acknowledges support from the European FP7 Marie Curie project PIEF-GA-2009-251904.

**Reference**

1. Castro, E. V. *et al.* Biased Bilayer Graphene: Semiconductor with a Gap Tunable by the Electric Field Effect. *Phys. Rev. Lett.* **99**, 216802, (2007).
2. Ohta, T., Bostwick, A., Seyller, T., Horn, K. & Rotenberg, E. Controlling the electronic structure of bilayer graphene. *Science* **313**, 951-954, (2006).
3. Zhang, Y. B. *et al.* Direct observation of a widely tunable bandgap in bilayer graphene. *Nature* **459**, 820, (2009).
4. Mak, K. F., Lui, C. H., Shan, J. & Heinz, T. F. Observation of an Electric-Field-Induced Band Gap in Bilayer Graphene by Infrared Spectroscopy. *Phys. Rev. Lett.* **102**, 256405, (2009).
5. Oostinga, J. B., Heersche, H. B., Liu, X. L., Morpurgo, A. F. & Vandersypen, L. M. K. Gate-induced insulating state in bilayer graphene devices. *Nat. Mater.* **7**, 151-157, (2008).
6. Xia, F. N., Farmer, D. B., Lin, Y. M. & Avouris, P. Graphene Field-Effect Transistors with High On/Off Current Ratio and Large Transport Band Gap at Room Temperature. *Nano Lett.* **10**, 715-718, (2010).
7. Ohta, T. *et al.* Interlayer interaction and electronic screening in multilayer graphene investigated with angle-resolved photoemission spectroscopy. *Phys. Rev. Lett.* **98**, (2007).
8. Mak, K. F., Sfeir, M. Y., Misewich, J. A. & Heinz, T. F. The evolution of electronic structure in few-layer graphene revealed by optical spectroscopy. *Proc. Natl. Acad. Sci. U. S. A.* **107**, 14999, (2010).
9. Mak, K. F., Shan, J. & Heinz, T. F. Electronic Structure of Few-Layer Graphene: Experimental Demonstration of Strong Dependence on Stacking Sequence. *Phys. Rev. Lett.* **104**, 176404, (2010).
10. Craciun, M. F. *et al.* Trilayer graphene is a semimetal with a gate-tunable band overlap. *Nat. Nanotechnol.* **4**, 383, (2009).
11. Bruna, M. & Borini, S. Observation of Raman G -band splitting in top-doped few-layer graphene. *Phys. Rev. B* **81**, 125421, (2010).
12. Jung, N. *et al.* Charge Transfer Chemical Doping of Few Layer Graphenes: Charge Distribution and Band Gap Formation. *Nano Lett.* **9**, 4133-4137, (2009).
13. Latil, S. & Henrard, L. Charge carriers in few-layer graphene films. *Phys. Rev. Lett.* **97**, (2006).
14. Aoki, M. & Amawashi, H. Dependence of band structures on stacking and field in layered graphene. *Solid State Commun.* **142**, 123, (2007).
15. Guinea, F., Neto, A. H. C. & Peres, N. M. R. Electronic states and Landau levels in graphene stacks. *Phys. Rev. B* **73**, 245426, (2006).
16. Min, H. K. & MacDonald, A. H. Electronic Structure of Multilayer Graphene. *Progress of Theoretical Physics Supplement* **176**, 227, (2008).
17. Norimatsu, W. & Kusunoki, M. Selective formation of ABC-stacked graphene layers on SiC(0001). *Phys. Rev. B* **81**, 161410, (2010).
18. Koshino, M. Interlayer screening effect in graphene multilayers with ABA and ABC stacking. *Phys. Rev. B* **81**, 125304, (2010).
19. Koshino, M. & McCann, E. Gate-induced interlayer asymmetry in ABA-stacked trilayer graphene. *Phys. Rev. B* **79**, 125443, (2009).
20. Koshino, M. & McCann, E. Trigonal warping and Berry's phase N pi in ABC-stacked multilayer graphene. *Phys. Rev. B* **80**, 165409, (2009).
21. Avetisyan, A. A., Partoens, B. & Peeters, F. M. Electric-field control of the band gap and Fermi energy in graphene multilayers by top and back gates. *Phys. Rev. B* **80**, 195401, (2009).
22. Avetisyan, A. A., Partoens, B. & Peeters, F. M. Electric field tuning of the band gap in graphene multilayers. *Phys. Rev. B* **79**, 035421, (2009).
23. Avetisyan, A. A., Partoens, B. & Peeters, F. M. Stacking order dependent electric field tuning of the band gap in graphene multilayers. *Phys. Rev. B* **81**, 115432, (2010).
24. Zhang, F., Sahu, B., Min, H. & MacDonald, A. H. Band structure of ABC -stacked graphene trilayers. *Phys. Rev. B* **82**, 035409, (2010).
25. Balog, R. *et al.* Bandgap opening in graphene induced by patterned hydrogen adsorption. *Nat Mater* **9**, 315-319, (2010).
26. Lui, C. H. *et al.* Imaging Stacking Order in Few-Layer Graphene. *Nano Lett.* **11**, 164-169, (2010).
27. Mak, K. F. *et al.* Measurement of the Optical Conductivity of Graphene. *Phys. Rev. Lett.* **101**, 196405, (2008).
28. Lu, C., Fu, Q., Huang, S. & Liu, J. Polymer Electrolyte-Gated Carbon Nanotube Field-Effect Transistor. *Nano Lett.* **4**, 623-627, (2004).
29. Li, Z. Q. *et al.* Band Structure Asymmetry of Bilayer Graphene Revealed by Infrared Spectroscopy. *Phys. Rev. Lett.* **102**, 4, (2009).

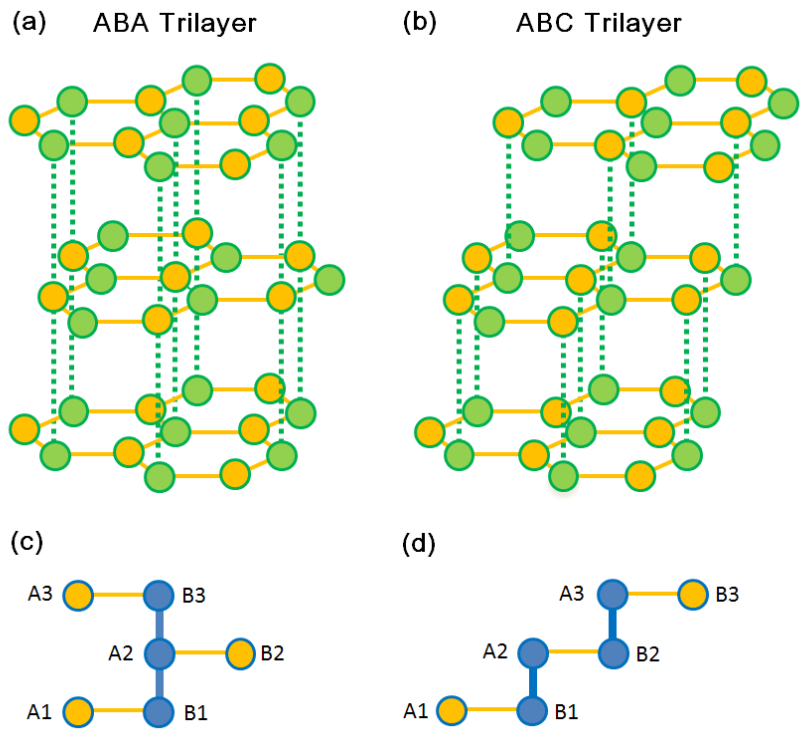

**Figure 1 | Crystal structure and tight-binding diagrams for trilayer graphene with ABA and ABC stacking order. a, b,** Crystal structure of ABA (a) and ABC (b) trilayer graphene. The yellow and green dots represent the A and B sublattices of the graphene honeycomb structure, respectively. **c, d,** Tight-binding diagrams for ABA (c) and ABC (d) trilayer graphene. At the $K$ point, the effective intralayer coupling vanishes. The atoms in yellow then become non-bonding monomers and the atoms in blue form a trimer in ABA trilayer and two dimers in ABC trilayer.

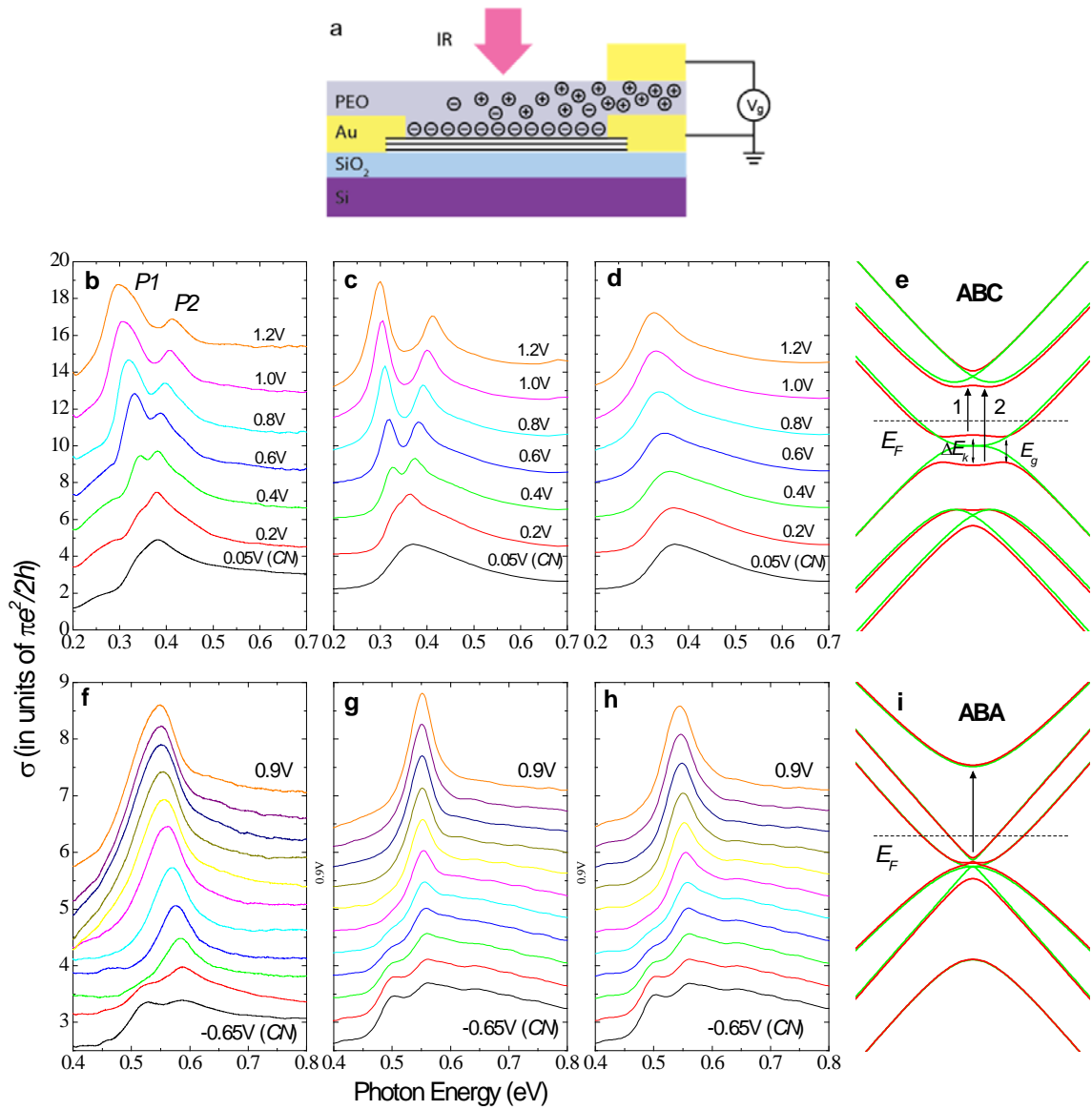

**Figure 2 | Comparison of optical conductivity $\sigma(\hbar\omega)$ of ABA and ABC graphene trilayers at different gate voltages $V_g$. a,** Schematic representation of the trilayer device used in these studies and described in the Methods section. **b,** Experimental gate dependent optical conductivity spectra $\sigma(\hbar\omega)$ of ABC-stacked trilayer graphene. **c,d,** Theoretical simulations of $\sigma(\hbar\omega)$ for ABC-stacked trilayer graphene under the same gating condition as in (b). (c) shows the predictions of TB model for the electronic structure described in the text, while (d) is a reference calculation in which the band structure is assumed to remain unaltered with gating and only the induced population changes are taken into account. In (b-d), the individual spectra are displaced by 2 units. The gate voltages $V_g$ and the condition of charge neutrality ($V_g = V_{CN} = -0.65$ V) are denoted on the spectra. **e,** The band structure of ABC trilayer graphene with (red) and

without (green) the presence of a perpendicular electric field as calculated within the TB model described in the text. Transitions 1 and 2 are the strongest optical transitions near the *K* point for electron doping. **f-h,** Results corresponding to (b-d) for ABA-stacked trilayer graphene samples. The different spectra, from top to bottom, were obtained for gate voltages $V_g$ = 0.9, 0.7, 0.5, 0.3, 0.1, -0.1, -0.3, -0.4, -0.5, -0.6, -0.65(*CN*) V and are displaced from one another by 0.4 units. **i,** Band structure of ABA trilayer graphene with (red) and without (green) the presence of a perpendicular electric field as calculated within the TB model described in the text. The arrow indicates the transition responsible for the main absorption peak in 0.5-0.6 eV.

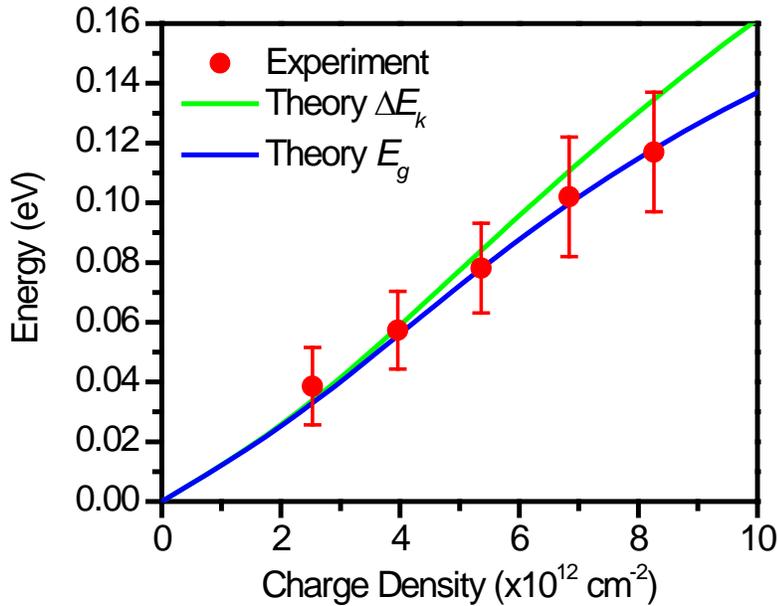

**Figure 3 | Dependence of the energy gap on the charge doping density of the ABC trilayer graphene.** The symbols are experimental data. The error bars arise primarily from uncertainties in determining the peak position of the absorption features. The results of TB model for both the gap at the *K*-point $\Delta E_K$ (green line) and the band gap $E_g$ (blue line) are plotted for comparison.

# Supplementary Information

## 1. Infrared conductivity in gated ABC-stacked graphene trilayers with hole doping

Figure S1 shows the conductivity $\sigma(\hbar\omega)$ of ABC-stacked graphene trilayer at gate biases corresponding to induced hole doping. The results are qualitatively the similar to those for the electron doping described in the main text. For hole doping, we observe an enhancement and splitting of the main transition peak at $\hbar\omega = 0.35$ eV. Just as for electron doping, this behavior is the result of the induction of the band gap. The high-energy component of the split peaks is broadened and reduced in strength at the highest gate bias voltages. We attribute this behavior to the lateral inhomogeneity in the electric-field of the polymer electrolyte top gate rather than to the inherent material response of the trilayer graphene.

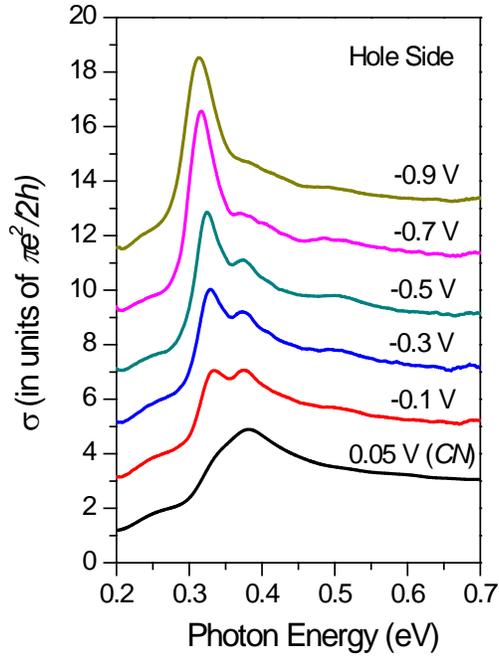

**Figure S1 Measured optical conductivity spectra $\sigma(\hbar\omega)$ of ABC-stacked graphene trilayer at gate biases corresponding to induced hole doping.** The spectra are displaced for 2 units for clarity. The gate voltage $V_g$ and charge neutrality (*CN*) point are denoted on the spectra. The corresponding spectra for electron doping are shown in figure 2b of the main text.

## 2. Electron-hole asymmetry in ABA-stacked graphene trilayer

In contrast to the result for the ABC-stacked trilayer structure, the ABA trilayer displays a clear difference in its optical conductivity $\sigma(\hbar\omega)$ for applied fields corresponding to electron and hole doping (Figure S2a). At the charge neutrality point $V_{CN} = -0.65$ V, $\sigma(\hbar\omega)$ exhibits two peaks, one at $\hbar\omega = 0.520$ eV and one at 0.585 eV. As we increase the bias $V_g$ (electron side), the amplitude of the higher-energy transition (0.585 eV) grows and the peak position red shifts, while the lower-energy transition (0.520 eV) subsides and disappears. As we decrease $V_g$ (hole side), the low-energy peak grows and the high-energy peak subsides and disappears. The evolution of the energy of the absorption peak with gate bias is summarized in figure S2d.

The observed behavior in the ABA-stacked trilayer can be understood within the framework of a tight-binding (TB) model that includes not only the dominant intralayer ($\gamma_0$) and interlayer ($\gamma_1$) couplings, but also the on-site energy difference $\delta$ and the parameter $v_4=\gamma_4/\gamma_0$ describing the next-nearest neighbor interlayer coupling. The observed electron-hole asymmetry is attributed to the band structure asymmetry between the valence and conduction bands, as previously discussed for graphene bilayer [S1-3]. According to the TB analysis of ABA trilayer (Figure 1c in the paper), sites A1, B2, A3 and B1, A2, B3 possess different energies ($\delta$) because of the different crystal field environment. This leads to different transition energies for the electron and hole sides. The next-nearest-neighbor interlayer coupling parameter $v_4$ produces different dispersion properties for the conduction and valence bands. This parameter is thus responsible for the different evolution of the electron and hole transition peaks with the gate voltage. The calculated band structure (Figure S2e) clearly shows the role of coupling parameters $\delta$ and $v_4$.

For a quantitative understanding on the ABA trilayer data, we have simulated the ABA trilayer IR conductivity by means of the Kubo formula with a 20-meV phenomenological broadening parameter. The conductivity at different induced charge densities $n$ is calculated in a self-consistent scheme that takes into account the different potentials at individual graphene layers resulted from uneven charge distribution in the sample (see Methods in the main text). We find that the main features of the experimental conductivity spectra and the dependence of absorption peak on the gate voltage are reproduced with $\gamma_1$=371 meV, $\delta$=37 meV and $v_4$=0.05 (Figure S2b and blue solid line in figure S2d). We have used a top-gate capacitance $C = 0.8$ μFcm$^{-2}$ to obtain the best description of the data. For comparison, we show the predicted $\sigma(\hbar\omega)$ spectra for a TB model of the same parameters, but under the neglect of any induced modification of the band structure (Figure S2c). The calculated spectra are quite similar to those found when the change of the band structure is considered (Figure S2b). The predicted absorption peaks are also in reasonable agreement with the experimental data (green dashed lines in Figure S2d). We conclude therefore that the gate-induced modification of band structure is not needed to explain our experimental data.

The extracted on-site energy difference ($\delta = 37$ meV) in ABA-stacked trilayer is much larger than the corresponding value in the AB-stacked bilayer ($\delta = 18$-25 meV) [S1-3]. The value is also much larger than the limit of $\delta < 22$ meV that we have estimated for the ABC stacked trilayer by considering the 44-meV width of the optical transition for $V_g = V_{CN}$ (figure 2b in the main text). The distinct behavior in the two cases is related to the difference in the crystal structure. Since $\delta$ arises from the change of local crystal field by the interlayer coupling, a trimer with two interlayer bonds is expected to have a larger value of $\delta$ than a dimer with only one interlayer bond. ABA trilayers, which feature trimers, should therefore exhibit a larger electron-hole asymmetry than do bilayers or ABC trilayers, which only have dimers.

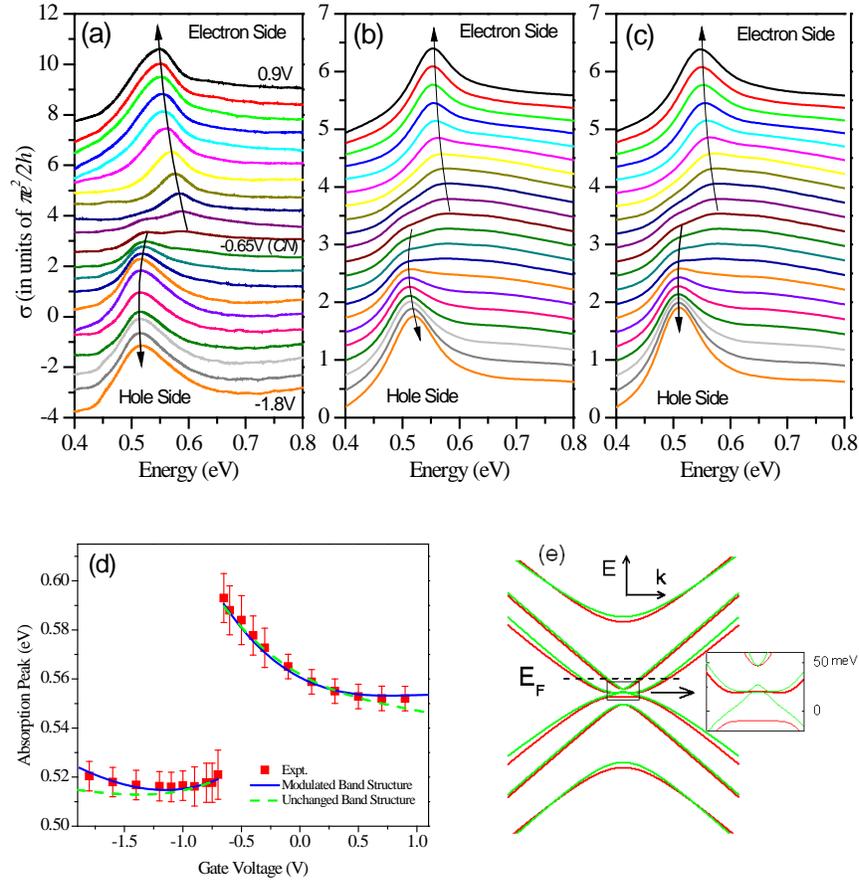

**Figure S2 Comparison of optical conductivity $\sigma(\hbar\omega)$ with theory for the gated ABA-stacked graphene trilayer for both electron and hole doping.** (**a**) Experimental conductivity spectra $\sigma(\hbar\omega)$ as a function of gate voltage $V_g$. The spectra are shifted for 0.6 units for clarity. The arrows are guides of peak positions. For charge neutrality $V_g = V_{CN} = -0.65$ V. From the top to bottom $V_g$ varies as 0.9, 0.7, 0.5, 0.3, 0.1, -0.1, -0.3, -0.4, -0.5, -0.6, -0.65(CN), -0.7, -0.75, -0.8, -0.9, -1.0, -1.1, -1.2, -1.4, -1.6 and -1.8 V. (**b,c**) Simulated spectra for $\sigma(\hbar\omega)$ from the Kubo formula under the same conditions. The spectra are shifted by 0.25 units for clarity. The results in (b) include the predicted modification of the band structure, while (c) is a reference calculation in which the band structure is assumed to remain unchanged. (**d**) Positions of absorption peaks extracted from (a) as a function of $V_g$. The solid blue and dashed green lines are from the calculated spectra in (b) and (c), respectively. (**e**) ABA trilayer graphene band structure with finite values (green) and zero values (red) for the TB parameters $\delta$ and $\gamma_4$.

**Reference**

S1   Li, Z. Q. *et al.* Band Structure Asymmetry of Bilayer Graphene Revealed by Infrared Spectroscopy. *Phys. Rev. Lett.* **102**, 4, (2009).

S2   Zhang, L. M. *et al.* Determination of the electronic structure of bilayer graphene from infrared spectroscopy. *Phys. Rev. B* **78**, 11, (2008).

S3   Mak, K. F., Lui, C. H., Shan, J. & Heinz, T. F. Observation of an Electric-Field-Induced Band Gap in Bilayer Graphene by Infrared Spectroscopy. *Phys. Rev. Lett.* **102**, 256405, (2009).